\documentclass[apj]{emulateapj}

\usepackage{mathptmx}
\usepackage{verbatim}
\usepackage{graphicx}
\usepackage{epstopdf}
\usepackage{amsmath}
\usepackage{textcomp}
\usepackage{latexsym}
\usepackage{multirow}
\usepackage{wasysym}

\newcommand{\ltsim}{\raisebox{-.5ex}{$\;\stackrel{<}{\sim}\;$}}
\newcommand{\gtsim}{\raisebox{-.5ex}{$\;\stackrel{>}{\sim}\;$}}
\newcommand{\kms}{\ifmmode {\rm km\ s}^{-1} \else km s$^{-1}$\fi}
\newcommand{\mbh}{$M_{\rm BH}$}
\newcommand{\lledd}{$L/L_{\rm Edd}$}

\newcommand{\xray}{\hbox{X-ray}}

\newcommand{\hb}{H$\beta$}
\newcommand{\civ}{C~{\sc iv}}

\journalinfo{The Astrophysical Journal, ???:??? (??pp), 2015 ??? ??}
\slugcomment{Received 2015 January 27; accepted 2015 March 17}

\shortauthors{SHEMMER AND LIEBER}
\shorttitle{WLQS AND THE MODIFIED BALDWIN EFFECT}

\begin{document}
\title{Weak Emission Line Quasars in the Context of a Modified Baldwin Effect}
\author{
Ohad~Shemmer\altaffilmark{1}
and Sara~Lieber\altaffilmark{1}
}
\altaffiltext{1}
		   {Department of Physics, University of
                      North Texas, Denton, TX 76203, USA; ohad@unt.edu}

\begin{abstract}

We investigate the relationship between the rest-frame equivalent width (EW) of the \ion{C}{4}\,$\lambda1549$ broad-emission line, monochromatic luminosity at rest-frame 5100\,\AA, and the \hb-based Eddington ratio in a sample of 99 ordinary quasars across the widest possible ranges of redshift ($0<z<3.5$) and bolometric luminosity ($10^{44}\ltsim L\ltsim10^{48}$ erg\,s$^{-1}$).
We find that EW(\ion{C}{4}) is primarily anti-correlated with
the Eddington ratio, a relation we refer to as a modified Baldwin effect (MBE), an extension of the result previously obtained for
quasars at $z<0.5$.
Based on the MBE, weak emission line quasars (WLQs), typically showing EW(\ion{C}{4})\ltsim10\,\AA, are expected to
have extremely high Eddington ratios.
By selecting all WLQs with archival \hb\ and \ion{C}{4} spectroscopic data, nine sources in total, we find that their \hb-based Eddington ratios are
typical of ordinary quasars with similar redshifts and luminosities.
Four of these WLQs can be accommodated by the MBE,
but the other five deviate significantly from this relation, at the $\gtsim3\sigma$ level, by exhibiting \ion{C}{4} lines much weaker than predicted from their \hb-based Eddington ratios.
Assuming the supermassive black-hole masses in all quasars can be determined reliably using the single-epoch \hb-method, our results indicate that EW(\ion{C}{4}) cannot depend solely on the Eddington ratio.
We briefly discuss a strategy for further investigation into the roles that basic
physical properties play in controlling the relative strengths of broad-emission lines in quasars.

\end{abstract}


\section{INTRODUCTION}
\label{sec:introduction}
The classical `Baldwin effect' is an anti-correlation between the rest-frame equivalent width (EW) of a broad-emission line region (BELR) line and
quasar luminosity, first observed for the \ion{C}{4}\,$\lambda1549$ line \citep{1977ApJ...214..679B}.
This anti-correlation is stronger and steeper for BELR lines with higher ionization potentials \citep[$\chi_{\rm ion}$;][]{2002ApJ...581..912D}, but it
involves substantial scatter, hampering its use as a cosmological probe \citep{1999ASPC..162..235O}. 
Considerable effort has been invested in attempts to minimize this scatter,
using partial-correlation and principal-component analyses involving emission-line as well as broad-band spectroscopic data \citep[e.g.,][]{1999ApJ...513...76W,2003ApJ...586...52S}, but the exact cause
of the Baldwin effect remains elusive.
A dependence on the shape of the continuum-source spectral energy distribution \citep[SED;][]{1993ApJ...415..517Z}, cosmic evolution \citep[][]{2001ApJ...556..727G}, or the supermassive black-hole mass \citep[\mbh;][]{2008MNRAS.389.1703X} being the primary physical driver for the EW-luminosity anti-correlation are among the explanations proposed for this effect.
It had also been speculated that the Baldwin effect depends largely on the normalized accretion rate, in terms of the Eddington ratio, \lledd,
where $L$ is the bolometric luminosity and $L_{\rm Edd}$ is the Eddington luminosity \citep[e.g.,][]{1999ASPC..162..395B,1999ApJ...515L..53W,2004ApJ...617..171B}.

Using the empirical BELR size-luminosity relation \citep{2005ApJ...629...61K,2009ApJ...697..160B} and assuming the BELR gas is virialized, \mbh\ takes a general expression of the form \mbh$\propto (\nu L_{\nu})^{0.5} {\rm FWHM}^2$; the Eddington ratio can therefore be expressed as \lledd$\propto (\nu L_{\nu})^{0.5} {\rm FWHM}^{-2}$, where $\nu L_{\nu}$ and FWHM typically correspond to the monochromatic luminosity at rest-frame 5100\,\AA\ and the full width at half maximum intensity of the broad \hb\footnote{Attempts to determine \mbh\ from high-ionization BELR lines, such as \civ, may yield unreliable results, since the line profiles are complicated by a non-virial (i.e., `wind') component \citep[e.g.,][]{2005MNRAS.356.1029B,2011AJ....141..167R,2012ApJ...753..125S,2012MNRAS.427.3081T}.} line, respectively \citep[see also, e.g.,][]{1998ApJ...505L..83L}.

Utilizing optical spectroscopic data for a sample of 81 quasars with \hbox{$L\sim 10^{44}-10^{46}$} erg\,s$^{-1}$ at $z<0.5$ from \cite{1992ApJS...80..109B}, and for which high-quality archival UV spectroscopic data were available, \citet[][hereafter BL04]{2004MNRAS.350L..31B}
found a significant anti-correlation between EW(\ion{C}{4}) and \hb-based \lledd;
they did not find a significant correlation between EW(\ion{C}{4}) and monochromatic luminosity at rest-frame 3000\,\AA.
BL04 argued that most of the scatter in the classical Baldwin effect is produced by a range of \lledd\ at a given $L$, driven by a range in
FWHM(\hb).
This scatter is minimized considerably when a combination of FWHM(\hb) and luminosity, i.e., the Eddington ratio,
is employed, thus strengthening the anti-correlation with EW(\ion{C}{4}).
BL04 claimed that the classical Baldwin effect is only a secondary effect since, typically, more luminous quasars also have higher Eddington ratios.
\citet[][]{2009ApJ...703L...1D} report a similar result for the EW of the \ion{Mg}{2}~$\lambda\lambda 2796, 2803$ doublet using \ion{Mg}{2}-based \lledd\ determinations for a sample of 2092 active galactic nuclei at $0.45 \leq z \leq 0.8$, suggesting that the Baldwin effect
is governed by \lledd.

In this work, we extend the BL04 analysis by including quasars with \hbox{$L\sim 10^{46}-10^{48}$} erg\,s$^{-1}$ at \hbox{$2 < z < 3.5$} that have \hb\ spectral information from near-infrared (NIR) spectroscopy as well as \ion{C}{4} information from optical spectroscopy in order to test whether the \hbox{EW(\ion{C}{4})-\lledd} anti-correlation, hereafter the modified Baldwin effect (MBE), remains strong across the widest possible ranges of redshift, luminosity, and \lledd.
In particular, the extension of this relationship to higher redshifts and luminosities is required in order to test the hypothesis that
the extreme weakness of the \ion{C}{4} lines in weak emission line quasars (WLQs), that typically have EW(\ion{C}{4})$\ltsim10$\,\AA\ (e.g., \citealt{1999ApJ...526L..57F}, \citealt{2009ApJ...699..782D}), is due to
extremely high accretion rates in these sources (see, e.g., \citealt{2007ApJS..173....1L}a; \citealt{2009ApJ...696..580S}, 2010).
In Section~\ref{sec:analysis} we describe the properties and spectroscopic measurements of our quasar sample, including WLQs,
and in Section~\ref{sec:results} we present the results of a correlation analysis involving EW(\ion{C}{4}), monochromatic luminosity, and \lledd.
In Section~\ref{sec:discussion} we discuss the implications of our results for quasars in general and for WLQs in particular,
and in Section~\ref{sec:conclusions} we summarize our main conclusions.
Throughout this paper, wavelengths, frequencies, and EWs are given in the rest-frame of each source.
Complete source names are given in Tables and Figures and abbreviated names
are given throughout the text.
Luminosity distances were computed using the standard cosmological model \citep[$\Omega_{\Lambda} = 0.7, \Omega_{\rm M} = 0.3, {\rm~and~}
H_0 = 70$~\kms~Mpc$^{-1}$; e.g.,][]{2007ApJS..170..377S}.

\section{Sample Selection and Data Analysis}
\label{sec:analysis}

Our high-redshift quasar sample is drawn from the \citet[][hereafter S04]{2004ApJ...614..547S} and \citet[][hereafter N07]{2007ApJ...671.1256N} studies involving high-quality NIR spectra of 29 and 15 sources, respectively, covering the \hb\ spectral region in the $2 < z < 3.5$ range.
We exclude six radio-loud quasars (RLQs) from S04,\footnote{S04 have, erroneously, identified \protect{[HB89]}~2132$+$014 as a RLQ, instead of \protect{[HB89]}~2126$-$158. They also identified \protect{[HB89]}~0329$-$385 and UM~645 as RLQs; however, as we mention below, these two sources have $10<R<100$ and are thus considered radio-intermediate quasars.} \protect{[HB89]}~0123$+$257, \protect{[HB89]}~0504$+$030, \protect{[HB89]}~2126$-$158, TON~618, UM~632, and \protect{[HB89]}~2254$+$024, as sources having radio-loudness values of $R>100$ (where $R$ is the ratio between the flux densities at 5\,GHz and 4400\,\AA; \citealt{1989AJ.....98.1195K}), based on the NRAO VLA Sky Survey (NVSS; \citealt{1998AJ....115.1693C}) for the first three of these sources and the Faint Images of the Radio Sky at Twenty Centimeters (FIRST) survey (\citealt{1995ApJ...450..559B}) for the latter three.
We also exclude \protect{[HB89]}~1246$-$057 (from S04) and SDSS~J2103$-$0600 (from N07) as broad-absorption line (BAL) quasars, based on \citet{1977ApJ...213..607O}  and \citet{2009ApJ...692..758G}, respectively.
The exclusion of RLQs and BAL quasars is intended to minimize potential effects of continuum boosting \citep[see below; e.g.,][]{2014A&A...568A.114M} and absorption biases (e.g., BL04), respectively, that may result in systematic underestimations of EW(\ion{C}{4}).

Relevant properties of our final sample of 36 `ordinary' quasars (i.e., type~1 quasars that are not radio loud and that do not have BALs) at high redshift, hereafter the HIZ sample, are given in Table~\ref{tab:HIZdata}.
We also note in Table~\ref{tab:HIZdata} that nine of the HIZ sources are identified as BAL quasars in \citet{2006ApJS..165....1T} but not in the more recent BAL quasar catalog of \citet{2009ApJ...692..758G}; we consider these sources as non-BAL quasars and they are retained in our sample.
For each source in the HIZ sample, we obtain the systemic redshift ($z_{\rm sys}$), $\nu L_{\nu} ( 5100$\,\AA$)$, and best-fit FWHM(\hb)
values from Tables~1~and~2 of S04 and from Table~2 of N07.
We derive the \lledd\ value for each source following Equation~(2) of \citet{2010ApJ...722L.152S},
\begin{equation}
L/L_{\rm Edd}=0.13 f(L) \left [ \frac{\nu L_{\nu}(5100\,{\rm \AA})}{10^{44}{\rm~erg~s}^{-1}} \right ]^{0.5} \left [ \frac{{\rm FWHM(H}\beta {\rm )}}{10^3 {\rm~km~s}^{-1}} \right ]^{-2},
\label{eq:Eddington}
\end{equation}
and using Equation~(21) of \citet{2004MNRAS.351..169M} to compute $f(L)$, the luminosity-dependent bolometric correction to $\nu L_{\nu} ( 5100$\,\AA$)$, which is in the range $5.42 < f(L) < 6.43$ for our sources.

Thirty of the HIZ sources have rest-frame UV spectra in electronic form that are publicly available; 23 spectra have been obtained from the Sloan Digital Sky Survey (SDSS; \citealt{2000AJ....120.1579Y}) and seven spectra have been obtained from the Two-Degree Field quasar redshift survey (2QZ; \citealt{2004MNRAS.349.1397C}); the spectral response of each 2QZ spectrum has been determined as described in S04.
For each spectrum, we fitted the $\sim1450$\,\AA$-1700$\,\AA\ spectral region around the \ion{C}{4} line using a linear continuum and two Gaussian profiles, describing the entire profile of the \ion{C}{4} line.
The two Gaussian profiles are used for least-squares fitting purposes only
and thus are not intended to represent two physically distinct emission regions.
The linear continuum was determined based on average flux densities obtained in 10\,\AA-wide intervals centered on \hbox{$\lambda_1 \simeq1445$\,\AA} and $\lambda_2 \simeq1695$\,\AA.
The EW of the \ion{C}{4} line in each source has been computed using the 
sum of the fluxes in each best-fit Gaussian profile and
the best-fit linear continuum underlying the emission line.
The errors on EW(\ion{C}{4}) were estimated by repeating the fitting procedure but, for each spectrum, the two steepest continua were considered, based on the $1\sigma$ value of the flux density in each of the two continuum intervals, i.e., fitting between
$(f_{\lambda_1} + \Delta f_{\lambda_1},  f_{\lambda_2} - \Delta f_{\lambda_2})$ and
$(f_{\lambda_1} - \Delta f_{\lambda_1},  f_{\lambda_2} + \Delta f_{\lambda_2})$.

For the six HIZ sources that lack publicly available spectra, we obtained the EW(\ion{C}{4}) values from the literature.
The EW(\ion{C}{4}) values for all of the HIZ sources are given in Table~\ref{tab:HIZdata}.
For the 23 sources from our HIZ sample that have SDSS spectra and for which we have measured EW(\ion{C}{4}) values,
such values (not shown in Table~\ref{tab:HIZdata}) can also be obtained from the spectral measurements of \citet{2011ApJS..194...45S}.
For 17 of these sources, the EW(\ion{C}{4}) values from \citet{2011ApJS..194...45S} agree with our measurements, within the errors.
The spectra of six sources for which the discrepancies between our measurements and the \citet{2011ApJS..194...45S} values
are \hbox{$\sim20\%-80$\%} have, on average, lower signal-to-noise ratios than the spectra of the 17 sources in which such discrepancies are
\hbox{$\ltsim20\%$}. Replacing our EW(\ion{C}{4}) measurements with the corresponding \citet{2011ApJS..194...45S} values for these six sources
does not alter significantly any of our subsequent results.

\begin{deluxetable*}{lcccccccc}
\tablecolumns{8}
\tabletypesize{\scriptsize}
\tablecaption{Basic Properties of the HIZ Sample}
\tablewidth{0pt}
\tablehead{
\colhead{Quasar} &
\colhead{$z_{\rm sys}$} &
\colhead{$\log \nu L_{\nu}(5100$\,\AA$)$} &
\colhead{FWHM(\hb)} &
\colhead{\lledd} &
\colhead{EW(\civ)} &
\colhead{Optical Ref.\tablenotemark{a}} &
\colhead{EW(\civ) Ref.\tablenotemark{d}} \\
\colhead{} &
\colhead{} &
\colhead{(erg s$^{-1}$)} &
\colhead{(\kms)} &
\colhead{} &
\colhead{(\AA)} &
\colhead{} &
\colhead{} \\
\colhead{(1)} &
\colhead{(2)} &
\colhead{(3)} &
\colhead{(4)} &
\colhead{(5)} &
\colhead{(6)} &
\colhead{(7)} &
\colhead{(8)}
}
\startdata
2QZ~J001221.1$-$283630      & 2.339 & 46.26 & 1915 & 2.82 & 32.3$^{+3.2}_{-2.4}$ & 1 & 2 \\
2QZ~J002830.4$-$281706      & 2.401 & 46.58 & 4833 & 0.63 & 39.8$^{+9.8}_{-7.3}$ & 1 & 2 \\
UM~667                                        & 3.132 & 46.28 & 3135 & 1.08 & 27.8$\pm$2.8 & 1 & 3 \\
LBQS~0109$+$0213                 & 2.349 & 46.80 & 5781 & 0.56 & $26.0^{+5.9}_{-3.9}$ & 1 & 4 \\
2QZ~J023805.8$-$274337      & 2.471 & 46.57 & 3437 & 1.22 & 25.8$^{+2.1}_{-1.3}$ & 1 & 2 \\
SDSS~J024933.42$-$083454.4\tablenotemark{b} & 2.491 & 46.38 & 5230 & 0.43 & 51.4$\pm$0.2 & 1 & 2, 5 \\
SDSS J025438.37$+$002132.8\tablenotemark{b} & 2.456 & 45.85 & 4164 & 0.38 & 66.6$^{+4.1}_{-2.6}$ & 6 & 2, 5 \\
\protect{[HB89]}~0329$-$385 & 2.435 & 46.71 & 7035 & 0.34 & 42.4$\pm$6.4\tablenotemark{c} & 1 & 7, 8 \\
SDSS J083630.55$+$062044.8 & 3.397 & 45.53 & 3950 & 0.30 & 14.9$^{+29.7}_{-6.4}$ & 6 & 2, 5 \\
SDSS J095141.33$+$013259.5\tablenotemark{b} & 2.411 & 45.55 & 4297 & 0.26 & 87.8$^{+5.9}_{- 5.3}$ & 6 & 2, 5 \\
SDSS~J100428.43$+$001825.6 & 3.046 & 46.44 & 3442 & 1.06 & 45.4$^{+2.9}_{-2.7}$ & 1 & 2, 5 \\
SDSS J100710.70$+$042119.1 & 2.363 & 45.17 & 5516 & 0.11 & 55.0$^{+16.0}_{-12.5}$ & 6 & 2, 5 \\
SDSS J101257.52$+$025933.2\tablenotemark{b} & 2.434 & 45.73 & 3892 & 0.39 & 34.9$^{+0.6}_{-0.1}$ & 6 & 2, 5 \\
SDSS J105511.99$+$020751.9 & 3.391 & 45.70 & 5424 & 0.19 & 49.9$^{+11.7}_{-9.3}$ & 6 & 2, 5 \\
SDSS J113838.26$-$020607.2 & 3.352 & 45.79 & 4562 & 0.30 & 26.1$^{+14.4}_{-5.0}$ & 6 & 2, 5 \\
SDSS J115111.20$+$034048.3\tablenotemark{b} & 2.337 & 45.58 & 5146 & 0.19 & 47.2$^{+2.3}_{-2.1}$ & 6 & 2, 5 \\
SDSS J115304.62$+$035951.5 & 3.426 & 46.04 & 5521 & 0.27 & 12.8$^{+6.7}_{-3.6}$ & 6 & 2, 5 \\
SDSS J115935.64$+$042420.0 & 3.451 & 45.92 & 5557 & 0.23 & 45.3$^{+4.9}_{-4.6}$ & 6 & 2, 5 \\
SDSS J125034.41$-$010510.5\tablenotemark{b} & 2.397 & 45.41 & 5149 & 0.16 & 72.3$\pm$0.2 & 6 & 2, 5 \\
\protect{[HB89]}~1318$-$113 & 2.306 & 46.89 & 4150 & 1.19 & 32.0$\pm$6.4 & 1 & 8 \\
\protect{[HB89]}~1346$-$036 & 2.370 & 46.88 & 5110 & 0.78 & 19.8$\pm$1.2\tablenotemark{c} & 1 & 7, 8 \\
SDSS~J135445.66$+$002050.2 & 2.531 & 46.49 & 2627 & 1.92 & 21.1$^{+2.0}_{-1.7}$ & 1 & 2, 5 \\
UM~629                                         & 2.460 & 46.56 & 2621 & 2.08 & 36.0$^{+3.6}_{-3.2}$ & 1 & 2, 5 \\
UM~642\tablenotemark{b}        & 2.361 & 46.29 & 3925 & 0.69 & 27.8$^{+2.3}_{-2.0}$ & 1 & 2, 5 \\
UM~645                                        & 2.257 & 46.31 & 3966 & 0.69 & 39.6$^{+9.3}_{-6.0}$ & 1 & 2, 5 \\
SBS~1425$+$606\tablenotemark{b} & 3.202 & 47.38 & 3144 & 3.55 & 44.7$^{+3.2}_{-6.2}$ & 1 & 2, 5 \\
SDSS J144245.66$-$024250.1 & 2.356 & 46.03 & 3661 & 0.60 & 53.7$^{+3.0}_{-3.3}$ & 6 & 2, 5 \\
SDSS J153725.36$-$014650.3 & 3.452 & 45.98 & 3656 & 0.57 & 34.5$^{+1.5}_{-1.4}$ & 6 & 2, 5 \\
SDSS~J170102.18$+$612301.0\tablenotemark{b} & 2.301 & 46.34 & 5760 & 0.34 & 18.7$^{+3.8}_{-3.2}$ & 1 & 2, 5 \\
SDSS~J173352.22$+$540030.5 & 3.428 & 47.00 & 3078 & 2.44 & 22.1$^{+16.0}_{-9.6}$ & 1 & 2, 5 \\
SDSS J210258.22$+$002023.4 & 3.328 & 45.79 & 7198 & 0.12 & 42.6$^{+7.9}_{-6.4}$ & 6 & 2, 5 \\
\protect{[HB89]}~2132$+$014 & 3.199 & 45.77 & 2505 & 0.98 & 36.4$\pm$3.6 & 1 & 9 \\
2QZ~J221814.4$-$300306      & 2.389 & 46.54 & 2986 & 1.57 & 47.4$^{+4.4}_{-4.0}$ & 1 & 2 \\
2QZ~J222006.7$-$280324      & 2.414 & 47.22 & 5238 & 1.07 & 20.7$\pm$1.5 & 1 & 2 \\
2QZ~J231456.8$-$280102      & 2.400 & 46.31 & 3459 & 0.91 & 73.2$^{+7.7}_{-7.0}$ & 1 & 2 \\
2QZ~J234510.3$-$293155      & 2.382 & 46.32 & 3908 & 0.72 & 47.6$^{+7.7}_{-4.1}$ & 1 & 2
\enddata
\tablerefs{(1) S04; (2) this work; (3) \citet{1993ApJ...415..563W}; (4) \citet{2001ApJS..134...35F}; (5) \citet{2011ApJS..194...45S};
(6) N07; (7) \citet{1989ApJ...342..666E}; (8) \citet{1977ApJ...213..607O}; (9) \citet{1991AJ....101.2004S}.}
\tablenotetext{a}{Source of rest-frame optical data, including $z_{\rm sys}$, $\nu L_{\nu}(5100$\,\AA$)$, and FWHM(\hb).}
\tablenotetext{b}{Identified as a BAL quasar in \citet{2006ApJS..165....1T} but not in \citet{2009ApJ...692..758G}.}
\tablenotetext{c}{The EW(\civ) value is the average of the two values given in the references; error bar is taken as one half the difference between the two values.}
\tablenotetext{d}{Unless stated otherwise, the EW(\civ) value adopted for analysis in this work is obtained from the first reference for each source.}
\label{tab:HIZdata}
\end{deluxetable*}

We complement the HIZ sample with a subset of 63 ordinary quasars from BL04, following
the exclusion of five BAL quasars, PG~0043$+$039, PG~2112$+$059 \citep{1998ApJS..118....1J}, PG~1001$+$054 \citep{2000ApJ...528..637B}, PG~1411$+$442 \citep{1987ApJ...322..729M}, and PG~1416$-$121 \citep{1986ApJ...310L...1T},
as well as 13 RLQs (with $R>100$), PG~0003$+$158, PG~0007$+$106, PG~1048$-$090, PG~1100$+$772, PG~1103$-$006, PG~1226$+$023, PG~1302$-$102, PG~1512$+$370, PG~1545$+$210, PG~1704$+$608, PG~2209$+$184, PG~2251$+$113, and PG~2308$+$098 \citep{1992ApJS...80..109B}.
For each of the 63 BL04 sources, we obtain the redshift and FWHM(\hb) information from Table~1 and Table~2 of \citet{1992ApJS...80..109B}, respectively, and EW(\ion{C}{4}) values are obtained from Table~1 of BL04.
The $\nu L_{\nu} (3000$\,\AA$)$ values for the BL04 sources, given in Table~1 of BL04, are converted to
$\nu L_{\nu} (5100$\,\AA$)$ values, assuming an optical continuum of the form $f_{\nu} \propto \nu^{-0.5}$ \citep[e.g.,][]{2001AJ....122..549V} and correcting the luminosity distances based on our adopted cosmological parameters (see Section~\ref{sec:introduction}).
The Eddington ratios of the BL04 sources are determined using Equation~\ref{eq:Eddington}.

\begin{deluxetable*}{lcccccccc}
\tablecolumns{8}
\tabletypesize{\scriptsize}
\tablecaption{Basic Properties of the WLQ Sample}
\tablewidth{0pt}
\tablehead{
\colhead{Quasar} &
\colhead{$z_{\rm sys}$} &
\colhead{$\log \nu L_{\nu}(5100$\,\AA$)$} &
\colhead{FWHM(\hb)} &
\colhead{\lledd} &
\colhead{EW(\civ)} &
\colhead{Optical Ref.\tablenotemark{a}} &
\colhead{EW(\civ) Ref.\tablenotemark{b}} \\
\colhead{} &
\colhead{} &
\colhead{(erg s$^{-1}$)} &
\colhead{(\kms)} &
\colhead{} &
\colhead{(\AA)} &
\colhead{} &
\colhead{} \\
\colhead{(1)} &
\colhead{(2)} &
\colhead{(3)} &
\colhead{(4)} &
\colhead{(5)} &
\colhead{(6)} &
\colhead{(7)} &
\colhead{(8)}
}
\startdata
SDSS~J083650.86$+$142539.0 & 1.749 & 45.93 & 2880 & 0.87 & $4.2^{+0.3}_{-0.5}$ & 1 & 1, 2 \\
SDSS~J094533.98$+$100950.1 & 1.683 & 46.17 & 4278 & 0.51 & $2.9^{+0.3}_{-0.6}$ & 1 & 1, 2 \\
SDSS~J114153.34$+$021924.3 & 3.55 & 46.55 & 5900 & 0.41 & $0.4\pm0.2$ & 3 & 4 \\
SDSS~J123743.08$+$630144.9 & 3.49 & 46.35 & 5200 & 0.42 & $7.7\pm1.1$ & 3 & 4, 2 \\
SDSS~J141141.96$+$140233.9 & 1.754 & 45.64 & 3966 & 0.34 & $3.8^{+0.8}_{-0.2}$ & 1 & 1, 2 \\
SDSS~J141730.92$+$073320.7 & 1.716 & 45.91 & 2784 & 0.92 & $2.5^{+2.1}_{-0.7}$ & 1 & 1, 2 \\
SDSS~J144741.76$-$020339.1 & 1.430 & 45.56 & 1923 & 1.33 & $7.7^{+0.2}_{-1.3}$ & 1 & 1 \\
SDSS~J152156.48$+$520238.5 & 2.238 & 47.14 & 5750 & 0.81 & $9.1\pm0.6$ & 5 & 5, 2 \\
PHL~1811                                       & 0.192 & 45.56 & 1943 & 1.30 & 6.6 & 6 & 6 
\enddata
\tablerefs{(1) \citet{2015ApJ....???..???P};
(2) \citet{2011ApJS..194...45S};
(3) \citet{2010ApJ...722L.152S};
(4) \citet{2009ApJ...699..782D};
(5) \citet{2011ApJ...736...28W};
(6) \citeauthor{2007ApJS..173....1L} (\citeyear{2007ApJS..173....1L}a).}
\tablenotetext{a}{Source of rest-frame optical data, including $z_{\rm sys}$, $\nu L_{\nu}(5100$\,\AA$)$, and FWHM(\hb).}
\tablenotetext{b}{The EW(\civ) value adopted for analysis in this work is obtained from the first reference for each source.}
\label{tab:WLQdata}
\end{deluxetable*}

In order to test the hypothesis that WLQs are quasars with extremely high Eddington ratios (e.g., \citealt{2009ApJ...696..580S}, 2010),
we select all the WLQs for which accurate \hb\ properties (such as FWHM and EW) are available from the literature.
For the purpose of this work, we consider all optically-selected type~1 quasars that i) have radio-loudness values of $R<100$, ii) do not show BAL troughs in their rest-frame UV spectra, and iii) have EW(\ion{C}{4})$<$10\,\AA\ as WLQs.
The third criterion follows from the fact that $\sim10$\,\AA\ marks the $3\sigma$ threshold on the low end of lognormal fits to distributions of
EW(\ion{C}{4}) values for quasars at $1.5\ltsim z\ltsim5.0$; i.e., $\ltsim0.15\%$ of quasars at this redshift range have EW(\ion{C}{4})$<10$\,\AA\ (e.g., \citealt{2009ApJ...699..782D}; \citealt{2011ApJ...736...28W}, 2012; \citealt{2015ApJ....???..???P}).\footnote{There is tentative evidence that the fraction of WLQs is considerably larger than 0.15\% of the entire quasar population at $z\gtsim5$ (see, e.g., \citealt{2006AJ....131.1203F}; \citealt{2014AJ....148...14B}).}
We caution that these selection criteria likely result in a heterogenous group of quasars, and we do not expect, a priori, a common origin for the weakness of the \ion{C}{4} BELR line in all such sources.
Our WLQ sample of nine sources includes
SDSS~J0836$+$1425, SDSS~J1411$+$1402, SDSS~J1417$+$0733, SDSS~J1447$-$0203 (\citealt{2010AJ....139..390P}, 2015),
SDSS~J0945$+$1009 (\citealt{2010MNRAS.404.2028H}; \citealt{2015ApJ....???..???P}),
SDSS~J1141$+$0219, SDSS~J1237$+$6301 \citep{{2009ApJ...699..782D},2010ApJ...722L.152S},
SDSS~J1521$+$5202 \citep{2007ApJ...665.1004J,2011ApJ...736...28W},  and PHL~1811
(\citealt{2007ApJS..173....1L}a, \citeyear{2007ApJ...663..103L}b).
Table~\ref{tab:WLQdata} presents the $z_{\rm sys}$, $\nu L_{\nu} (5100$\,\AA$)$, FWHM(\hb), \lledd\ (determined using Equation~\ref{eq:Eddington}), and EW(\ion{C}{4}) values for our WLQ sample.
For five SDSS sources from the WLQ sample, the EW(\ion{C}{4}) values from either \citet{2009ApJ...699..782D} or \citet{2015ApJ....???..???P}
are consistent, within the errors, with the values obtained from \citet{2011ApJS..194...45S}; \citet{2011ApJS..194...45S} do not provide EW(\ion{C}{4}) measurements for SDSS~J1141$+$0219 and SDSS~J1447$-$0203.
For SDSS~J1521$+$5202, \citet{2011ApJS..194...45S} give EW(\ion{C}{4})=$3.0\pm0.2$, which is a factor of $\simeq3$ smaller than the value reported in \citet{2011ApJ...736...28W}; see Table~\ref{tab:WLQdata}.

Finally, we note that our adoption of $R=100$ as the radio-loudness threshold, instead of the conventional (and more conservative)
threshold of $R=10$ (e.g., \citealt{1989AJ.....98.1195K}), is intended to exclude only sources that are more representative of the RLQ population
\citep{2002AJ....124.2364I}, for which the potential effects of continuum boosting are expected to be more pronounced. Our HIZ, BL04, and WLQ samples include four sources (\protect{[HB89]}~0329$-$385, UM~645, SDSS~J1733$+$5400, and SDSS J2102$+$0020), three sources
(PG~1211$+$143, PG~1309$+$355, and PG~1425$+$267),  and one source (SDSS~J1141$+$0219), respectively, with $10<R<100$.

\section{Results}
\label{sec:results}

\begin{figure*}
\epsscale{1.0}
\plotone{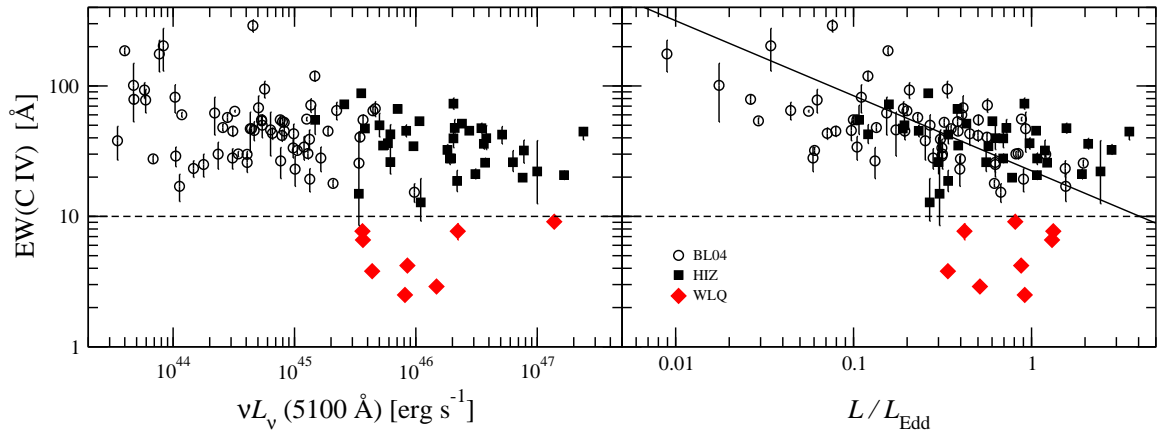}
\caption{EW(\civ) versus monochromatic luminosity at 5100\,\AA\ ({\it left}) and \lledd\ ({\it right}). Circles, squares, and diamonds represent the BL04, HIZ, and WLQ samples, respectively. The dashed line in each panel marks the EW(\civ)$=10$\,\AA\ threshold for WLQs, and the solid line in the right panel marks the BCES Bisector best-fit $\log$~[EW(\civ)]-$\log$~(\lledd) relation for a combination of the BL04 and HIZ samples. The WLQ SDSS~J114153.34$+$021924.3 with \lledd$=0.41$ and EW(\civ)$=0.4\pm0.2$ is not plotted, for clarity.}
\label{fig:L_L_LEdd_EW_C4}
\end{figure*}

We plot EW(\ion{C}{4}) versus $\nu L_{\nu} (5100$\,\AA$)$ and \lledd\ for the BL04 and HIZ samples in Fig.~\ref{fig:L_L_LEdd_EW_C4},
and present the respective Spearman-rank correlation coefficients ($r_{\rm S}$) and chance probabilities ($p$)
in Table~\ref{tab:Spearman}.
Our results for the BL04 sample indicate that EW(\ion{C}{4}) and \lledd\ are significantly anti-correlated (i.e., $p<1\%$), whereas
no significant correlation is observed between EW(\ion{C}{4}) and $\nu L_{\nu} (5100$\,\AA$)$, consistent with the BL04 finding.
For the HIZ sample, there is no significant correlation between EW(\ion{C}{4}) and either \lledd\ or $\nu L_{\nu} (5100$\,\AA$)$.
The lack of an EW(\ion{C}{4})-$\nu L_{\nu} (5100$\,\AA$)$ anti-correlation in our HIZ sample, as might have been expected from the classical Baldwin effect, may be due to obtaining \ion{C}{4} and $\nu L_{\nu} (5100$\,\AA$)$ from two different and non-contemporaneous spectra for each source, as
well as including different sources for \ion{C}{4} data with different measurement techniques (see Table~\ref{tab:HIZdata}).
We do find, however, a significant anti-correlation between EW(\ion{C}{4}) and $\nu L_{\nu} (1450$\,\AA$)$ for the HIZ sources, consistent with
the \citet{1977ApJ...214..679B} result.
When the BL04 and HIZ samples are combined, both $\nu L_{\nu} (5100$\,\AA$)$ and \lledd\  are significantly anti-correlated with EW(\ion{C}{4}), although the anti-correlation with \lledd\ is substantially stronger and it is stronger than the EW(\ion{C}{4})-\lledd\ anti-correlation for the BL04 sample alone ($p$ drops from \hbox{$2.11 \times 10^{-6}$} to \hbox{$3.03 \times 10^{-8}$}), thus bolstering the BL04 results.
We also note that, when replacing either $\nu L_{\nu} (5100$\,\AA$)$ or \lledd\ by source redshift, the above correlations with EW(\ion{C}{4}) weaken considerably.
These results indicate that the MBE is more pronounced at lower luminosities and thus lower redshifts, where several
low-luminosity sources with relatively high Eddington ratios are observed (Fig.~\ref{fig:L_L_LEdd_EW_C4}).
At high redshift, it is difficult to obtain high-quality spectral information for low-luminosity quasars.
This practical limitation results in a strong dependence between $L$ and \lledd, and thus
relatively high-$L$ sources have narrow ranges of both $L$ and \lledd, which may also explain why we do not detect a significant EW(\ion{C}{4})-\lledd\ anti-correlation for the HIZ sample alone.
In fact, the BL04 sample spans the 0.01\ltsim \lledd\ltsim 1 range, while the HIZ sample spans only the 0.1\ltsim \lledd\ltsim 1 range.

Fig.~\ref{fig:L_L_LEdd_EW_C4} also shows that WLQs, not included in any of the correlations, appear as outliers in these relations.
To quantify the deviation of WLQs from the EW(\ion{C}{4})-\lledd\ anti-correlation, we fitted a linear model to the $\log$~[EW(\ion{C}{4})] and
$\log$~(\lledd) values of sources from the combined BL04 and HIZ samples.
A standard $\chi^2$ minimization, weighted by the errors on $\log$~[EW(\ion{C}{4})], yielded an unsatisfactory fit (with $\chi^2 / \nu = 5191 / 97$),
indicating that either a linear model does not provide the best fit, the error bars on $\log$~[EW(\ion{C}{4})] are underestimated, or that additional scatter in the data must be taken into account.
Assuming a linear model with \hbox{$\chi^2 / \nu = 97 / 97$}, we find an additional scatter in the $\log$~[EW(\ion{C}{4})] values of $\sim0.2$ dex \citep[see, e.g.,][]{2002ApJ...574..740T,2005ApJ...629...61K}; this scatter is much larger than the typical measurement errors on $\log$~[EW(\ion{C}{4})].
One likely source for this scatter stems from the fact that the \ion{C}{4} and \hb\ spectral information are obtained from different datasets and are non-contemporaneous. But as we discuss in Section~\ref{sec:discussion}, additional physical parameters may also contribute to this scatter.
We account for this potential intrinsic scatter by using the bivariate correlated errors and scatter
method \citep[BCES;][]{1996ApJ...470..706A} for performing the linear regression.
Since a derivation of the Eddington ratio involves a typical uncertainty of $\sim0.3$ dex, we assign to all the $\log$~(\lledd) values of the BL04 and HIZ samples homoscedastic errors of 0.3~dex \citep[cf. Section 3.2.2 of][]{2008ApJ...682...81S}. The BCES Bisector best-fit relation for the BL04 and HIZ samples,
\begin{equation}
\log\,{\rm [EW}({\rm C~{\textsc i}{}\textsc v})] =   (-0.58\pm0.07) \log\,(L/L_{\rm Edd}) + (1.35\pm0.04),
\label{eq:BCES}
\end{equation}
is plotted in the right panel of Fig.~\ref{fig:L_L_LEdd_EW_C4}.
We cross-checked the above BCES Bisector relation against the results from a linear-regression analysis using the maximum-likelihood estimate method of \cite{2007ApJ...665.1489K}.
This method results in a flatter slope (-0.41$\pm$0.08) and a roughly similar intercept (1.44$\pm$0.04), but the slope is highly sensitive to the uncertainties assumed on $\log$~(\lledd) in the sense that the best-fit relation steepens as the errors increase beyond 0.3~dex.
We adopt the more conservative BCES Bisector relation since, as shown below, this provides more stringent constraints on the WLQ sample.

\begin{figure}
\epsscale{1.0}
\plotone{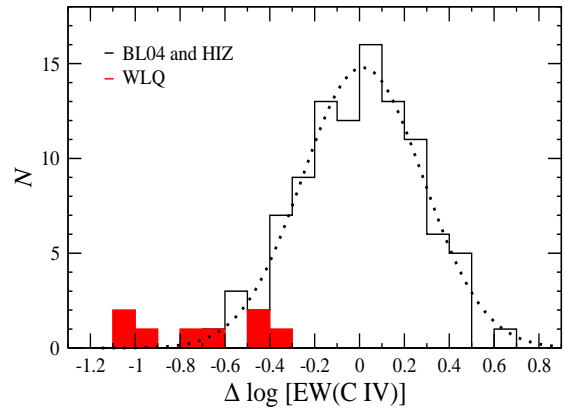}
\caption{Distribution of $\log$\,[EW(\civ)] residuals, computed as the difference between observed $\log$~[EW(\civ)] values and $\log$~[EW(\civ)] values predicted from the BCES Bisector best-fit $\log$~[EW(\civ)]-$\log$~(\lledd) relation. Sources from the BL04 and HIZ samples are represented by the unshaded histogram; the dotted curve is the best-fit Gaussian distribution for this
histogram with $\sigma=0.27$ dex.
Five of the nine WLQs (shaded histogram) lie below the $\sim3\sigma$ threshold of the best-fit Gaussian distribution
(including the WLQ SDSS~J114153.34$+$021924.3 with \hbox{$\Delta \log$\,[EW(\civ)]$\simeq-2.0$} which is not included in the shaded histogram, for clarity).}
\label{fig:Delta_log10_EW_C4}
\end{figure}

In Fig.~\ref{fig:Delta_log10_EW_C4} we plot a distribution of the differences between the observed $\log$\,[EW(\ion{C}{4})] values and those predicted from the \lledd\ values of the sources, based on Equation~\ref{eq:BCES}.
The distribution of these residuals, $\Delta \log$\,[EW(\ion{C}{4})], for the BL04 and HIZ samples is roughly symmetric with zero mean
and extreme values of $\pm0.6$~dex.
The best-fit Gaussian model to this distribution gives $\mu = -0.04$~dex and $\sigma = 0.27$~dex.
All sources from the WLQ sample lie at $\gtsim1.5\sigma$ below the mean of this Gaussian distribution and five of these lie at $\gtsim3\sigma$ (Fig.~\ref{fig:Delta_log10_EW_C4}).
The three WLQs with the largest (less negative) residuals ($-0.46\leq\Delta \log$\,[EW(\ion{C}{4})]$\leq-0.39$~dex), PHL~1811, SDSS~J1521$+$5202, and SDSS~J1447$-$0203, overlap with the residuals of the combined BL04 and HIZ samples (although SDSS~J1521$+$5202 will lie below the $3\sigma$ threshold if we adopt the \citeauthor{2011ApJS..194...45S} \citeyear{2011ApJS..194...45S} EW measurement); we discuss these sources further in Section~\ref{sec:discussion}.

\section{Discussion}
\label{sec:discussion}

In this work, we expanded the BL04 parameter space by including sources having the highest possible redshifts and luminosities, for which high-quality spectral information for \ion{C}{4} and \hb\ is available, in order to test whether the relative strength of \ion{C}{4} depends primarily on \lledd.
We find that, for ordinary quasars across the 10$^{43}$\ltsim$\nu L_{\nu} (5100$\,\AA$)$\ltsim 10$^{47}$\,erg\,s$^{-1}$ and 0.01\ltsim \lledd\ltsim1 ranges, the scatter in the Baldwin effect is minimized when \lledd\ replaces monochromatic luminosity, thus extending the EW(\ion{C}{4})-\lledd\ anti-correlation from BL04, i.e., the MBE, by almost two orders of magnitude in luminosity.
However, we also find no significant correlations between EW(\ion{C}{4}) and either $\nu L_{\nu} (5100$\,\AA$)$ or \lledd\ when only high-redshift and high-luminosity sources are considered; this is mainly a consequence of additional scatter introduced by using diverse data sets
and the strong dependence between $\nu L_{\nu} (5100$\,\AA$)$ and \lledd\ at high redshift.
We also investigate how WLQs fit into this picture and whether they have exceptionally high Eddington ratios.
We find that, in general, the \hb-based Eddington ratios of WLQs are within the norm when compared to ordinary quasars with similar redshifts and luminosities (see, e.g., Tables~\ref{tab:HIZdata}~and~\ref{tab:WLQdata}), and that most WLQs deviate considerably from the EW(\ion{C}{4})-\lledd\ anti-correlation, suggesting that the MBE may not be applicable to all quasars.
If the strong deviation of these WLQs is due to selection effects, then low-\lledd\ sources with \hbox{EW(\ion{C}{4})$\,\approx10^2$\,\AA$-10^3$~\AA} are required in order to cause the necessary steepening in the MBE for accommodating additional WLQs.
It will be interesting to see whether the emerging population of high-EW(\ion{C}{4}) quasars at high redshift would
produce such an effect \citep[e.g.,][]{2014arXiv1405.1047R}.

The fact that most WLQs do not follow the MBE may, instead, bring into question the reliability of determining \mbh\ values in WLQs
and perhaps in ordinary quasars as well.
Our linear regression analysis already indicates a trend of a steeper best-fit EW(\ion{C}{4})-\lledd\ relation as the assumed uncertainties on \lledd\ increase.
Such a steepening may accommodate some, but perhaps not all, WLQs in the MBE.
The standard, single-epoch \hb-method for obtaining \mbh\ and \lledd, briefly outlined in Section~\ref{sec:introduction} and given in Equation~\ref{eq:Eddington}, respectively, is likely too simplistic, and may involve uncertainties much larger than 0.3~dex (as we assume in Section~\ref{sec:results}) as well as systematic uncertainties.
One such systematic uncertainty may be a consequence of orientation bias \citep[see, e.g.,][]{2014Natur.513..210S}.
In this scenario, sources viewed close to pole-on exhibit narrower BELR lines, and thus their actual \mbh\ (\lledd) values should be higher (lower).
If WLQs suffer from orientation bias, then their Eddington ratios should be even smaller than those in Table~\ref{tab:WLQdata}, resulting in a larger deviation from the MBE.
Orientation bias is, therefore, an unlikely explanation for this deviation.
A different method of determining Eddington ratios in WLQs is required to test whether these ratios are considerably larger than the respective
\hb-based values.
The hard-\xray\ photon index ($\Gamma$) is one such \lledd\ indicator that can be used for cross-checking with \hb-based values \citep[e.g.,][]{2008ApJ...682...81S}.
To this end, such a comparison has been made for two WLQs, PHL~1811 and SDSS~J1141$+$0219; for both sources the \xray-based \lledd\ value is consistent with the \hb-based value (see, \citealt{2007ApJ...663..103L}b and \citealt{2010ApJ...722L.152S}, respectively).
Based on Equation~\ref{eq:BCES}, WLQs are expected to have \lledd\gtsim4, which would render extremely steep hard-\xray\ spectra with $\Gamma\gtsim3$ \citep{2008ApJ...682...81S}.
\xray\ spectroscopy of a statistically meaningful sample of WLQs may therefore provide a robust test of the hypothesis that WLQs are sources with extremely high Eddington ratios.

\begin{deluxetable}{llccc}
\tablecolumns{5}
\tabletypesize{\scriptsize}
\tablecaption{Spearman-Rank Correlation Coefficients}
\tablewidth{0pt}
\tablehead{
\colhead{Correlation} &
\colhead{Sample} &
\colhead{$N$} &
\colhead{$r_{\rm S}$} &
\colhead{$p$}
}
\startdata
EW(\civ)-$\nu L_{\nu}$ (5100 \AA) & BL04                  & 63 & $-0.23$ & $7.37 \times 10^{-2}$ \\
EW(\civ)-\lledd                                    & BL04                 & 63 & $-0.56$ & $2.11 \times 10^{-6}$ \\
EW(\civ)-$\nu L_{\nu}$ (5100 \AA) & HIZ                     & 36 & $-0.38$ & $2.32 \times 10^{-2}$ \\
EW(\civ)-\lledd                                    & HIZ                     & 36 & $-0.30$ & $7.74 \times 10^{-2}$ \\
EW(\civ)-$\nu L_{\nu}$ (5100 \AA) & BL04 and HIZ  & 99 & $-0.33$ & $8.35 \times 10^{-4}$ \\
EW(\civ)-\lledd                                    & BL04 and HIZ  & 99 & $-0.52$ & $3.03 \times 10^{-8}$
\enddata
\tablecomments{The last three columns represent the number of sources in each correlation, the Spearman-rank correlation coefficient, and the chance probability, respectively.}
\label{tab:Spearman}
\end{deluxetable}

Alternatively, WLQs may be pointing to the fact that additional physical properties may play a role in determining the relative strength
of the \ion{C}{4} line.
From a chronological perspective, the classical Baldwin effect \citep{1977ApJ...214..679B}, observed for high-redshift quasars (for practical reasons), included substantial scatter which, as we explain in Section~\ref{sec:results}, could not have been effectively minimized by replacing the luminosity  with the Eddington ratio.
Mainly low-luminosity sources with high Eddington ratios, e.g., narrow-line Seyfert 1 galaxies (NLS1s), led BL04 to conclude that \lledd\ is the primary physical parameter governing the relative strength of \ion{C}{4}.
NLS1s deviate considerably from the classical Baldwin effect, but are accommodated by the MBE.
In this work, we show that most of our WLQs deviate considerably from the MBE (and from the classical Baldwin effect), revealing that the
relative strength of \ion{C}{4} may not depend solely on \lledd\ for all quasars.
In this respect, WLQs are analogous to NLS1s by calling for more scrutiny into the parameters controlling BELR line strengths in quasars.

BL04 explored additional observables that may further reduce the scatter in the MBE.
For example, they found that the combination of \lledd\ and the EW of the [\ion{O}{3}]\,$\lambda5007$ narrow emission line provided the strongest anti-correlation with EW(\ion{C}{4}).
Only one of our WLQs, SDSS~J1447$-$0203, has [\ion{O}{3}] emission lines tentatively detected; the spectra of the other four WLQs from \citet{2015ApJ....???..???P} do not cover the [\ion{O}{3}] lines. Our other four WLQs, SDSS~J1141$+$0219, SDSS~J1237$+$6301, SDSS~J1521$+$5202, and PHL~1811 (as well as about a quarter of the sources from the HIZ sample; see \citealt{2004ApJ...614..558N}; N07) have tight upper limits on EW([\ion{O}{3}]).
While this may be consistent with the general trend of weaker \ion{C}{4} lines in sources with weaker [\ion{O}{3}] emission (see BL04 and references therein), the limited [\ion{O}{3}] statistics prevent us from testing whether this observable can explain part or all of the WLQ deviation.
We note, however, that the relative strength of [\ion{O}{3}] as well as other observables studied by BL04 may all be governed primarily by the Eddington ratio.

Additional parameters that may affect the relative strength of the \ion{C}{4} line can be split broadly into properties of the i) SED,
and ii) BELR.
A high Eddington ratio results in a softer, UV-peaked SED, and this may naturally explain relatively weak \ion{C}{4} lines
due to the paucity of highly ionizing photons; this model has been suggested for explaining the unusual properties of PHL~1811
(e.g., \citeauthor{2007ApJS..173....1L} \citeyear{2007ApJS..173....1L}a).
It is interesting to note that PHL~1811, its high-redshift `analog', SDSS~J1521$+$5202 \citep{2011ApJ...736...28W}, as well as 
SDSS~J1447$-$0203 which \citet{2015ApJ....???..???P} consider a `borderline' WLQ (or, an extreme `wind-dominated' quasar), appear to follow the MBE (within $\sim1.5\sigma-2\sigma$; see Section~\ref{sec:results}).
These sources may be different than the rest of the WLQs in our sample in the sense that the Eddington ratio alone may explain
their weak \ion{C}{4} lines.\footnote{Although this does not necessarily imply that these three sources belong to a single quasar subclass.
In particular, they differ in their \xray\ properties; SDSS~J1447$-$0203 and SDSS~J1521$+$5202 exhibit an effective power-law photon index
($\Gamma$) of $>1.0$ and $0.6\pm0.2$, respectively, in the observed-frame $0.5-8$~keV band, indicating significant intrinsic absorption at least in the latter source \citep{2015ApJ....???..???L}, and PHL~1811 exhibits  $\Gamma=2.3\pm0.1$ in the observed-frame 0.3-5~keV band with no detectable intrinsic absorption (\citealt{2007ApJ...663..103L}b).}
Alternatively, the difference between PHL~1811-like sources and the other, more extreme WLQs may be related to SED shielding (or modification) and orientation effects \citep{2011ApJ...736...28W}.
We emphasize that the EW(\ion{C}{4})$<10$\,\AA\ criterion we adopt for WLQs is statistically driven and it depends on the particular quasar sample
under consideration (see Section~\ref{sec:analysis}).
It is more instructive, perhaps, to use a physically-motivated definition for WLQs as being clear outliers from the MBE, deviating by more than $3\sigma$ from this relation on the low-EW end; i.e., sources having \hbox{$\Delta \log$\,[EW(\ion{C}{4})]$\ltsim-0.8$}, based on this work.
Given this definition, only five sources in our sample (i.e., further excluding SDSS~J1237$+$6301 with \hbox{$\Delta \log$\,[EW(\ion{C}{4})]$\sim-0.7$}) can be considered as WLQs, i.e., sources for which the \hb-based \lledd\ value may not fully explain their \ion{C}{4} line weakness.
Finally, we note that a `cold' accretion disk, due to high \mbh\ values, has also been offered to explain
the weak \ion{C}{4} lines in WLQs \citep{2011MNRAS.417..681L}; detailed UV spectroscopy of WLQs is required
to test the predictions of this model.

A variety of BELR physical properties can also affect the relative strength of the \ion{C}{4} line, such as the BELR geometry, covering factor,
density, and metallicity.
Extremely weak \ion{C}{4} lines, such as those observed in WLQs with \hbox{$\Delta \log$\,[EW(\ion{C}{4})]$\ltsim-0.8$}, may be attributed to a deficiency of gas in the BELR \citep[i.e., an `anemic' BELR;][]{2010ApJ...722L.152S}, or to an early evolutionary stage in the quasar's duty cycle where the BELR just started to form \citep[e.g.,][]{2010MNRAS.404.2028H}.
A more rigorous investigation of the parameters controlling the relative strengths of BELR lines in quasars, which is beyond the scope of this work, should include a comprehensive analysis of spectral information for low- and high-ionization BELR lines as well as the SED shape, in conjunction
with photoionization modeling, for a quasar sample much larger than studied herein.
Most importantly, the relative strengths of high-ionization BELR lines, such as \ion{C}{4} with \hbox{$\chi_{\rm ion}=47.9$}\,eV, should be investigated jointly with the relative strengths of low-ionization BELR lines, such as \hb\ with \hbox{$\chi_{\rm ion}=13.6$}\,eV or \ion{Mg}{2} with
\hbox{$\chi_{\rm ion}=7.6$}\,eV.
For example, correlations involving ratios of the relative strengths of these lines, such as EW(\ion{C}{4})/EW(\hb), as well as the \xray-to-optical SED should be investigated in more detail (e.g., BL04; \citealt{2011ApJ...736...28W}, 2012; \citealt{2015ApJ....???..???P}).
Furthermore, it is necessary to decompose the BELR lines into `disk' and `wind' (or outflow) components, in particular for \ion{C}{4} \citep[e.g.,][]{2011AJ....141..167R}, in order to check whether the EW of each component of the line profile
is correlated with a fundamental physical property, such as \lledd.
Detailed line-profile measurements, yielding emission-line blueshifts and line asymmetries, should provide additional insights
\citep[e.g.,][]{2011AJ....141..167R}.

\section{Conclusions}
\label{sec:conclusions}

We utilize a sample of 99 ordinary quasars across wide ranges of luminosity and redshift to show that the relative strength of the broad \ion{C}{4} line
is primarily anti-correlated with the \hb-based Eddington ratio, i.e., a
MBE, thus confirming and extending previous work limited to nearby, low-luminosity sources.
We also find that all nine WLQs with available \hb\ and \ion{C}{4} information in the archive have typical \hb-based \lledd\ values
in contrast with the extremely high values expected from the MBE.
While the EWs of the \ion{C}{4} lines in four of these WLQs are consistent with the MBE, the other five WLQs deviate significantly
from this relation by exhibiting EWs much smaller than predicted from their \hb-based \lledd\ values.
In case the single-epoch \hb-method can provide a reliable determination of \mbh\ in all quasars, then our results indicate that EW(\ion{C}{4}) cannot depend solely on \lledd.
While a comprehensive investigation into the nature of the MBE is beyond the scope of this study,
we outline additional spectroscopic work required to determine
the roles that basic quasar physical properties play in controlling the relative strengths of broad-emission lines in quasars.

\acknowledgements

We thank Marcia Lieber and Richard Plotkin for fruitful discussions.
We gratefully acknowledge a helpful and constructive report from an anonymous referee who helped to improve this work.
This research has made use of the NASA/IPAC Extragalactic Database (NED) which is operated by the Jet Propulsion Laboratory, California Institute of Technology, under contract with the National Aeronautics and Space Administration.

\end{document}